\documentclass[a4paper,10pt]{article}
\usepackage{amsmath,amsfonts}
\usepackage{graphicx}
\usepackage{hyperref}
\usepackage[title]{appendix}
\newcommand{\dd}{\textrm{d}}

\begin{document}
\title{\bf Connection probabilities in the double-dimer model -- the case of two connectivity patterns}
\author{Nahid Ghodratipour\footnote{ghodratipour\_n@physics.sharif.edu}\quad and\quad Shahin Rouhani\footnote{srouhani@sharif.ir} \\
	\small Department of Physics, Sharif University of Technology, \\
    \small and School of Particles and Accelerators, IPM, \\ \small Tehran, Iran}
\maketitle

%=========================================================================%
\begin{abstract}
We apply the Grassmannian representation of the dimer model, an equivalent approach to Kasteleyn's solution to the close-packed dimer problem, to calculate the connection probabilities for the double-dimer model with wired/free/wired/free boundary conditions, on a rectangular subdomain of the square lattice with four marked boundary points at the corners. Using some series identities related to Schwarz-Christoffel transformations, we show that the continuum of the result is consistent with the corresponding one in the upper half-plane (previously obtained by Kenyon-Wilson), which is in turn identical to the connection probabilities for 4SLE$_4$ emanating from the boundary, or equivalently, to a conditioned version of CLE$_4$ with wired/free/wired/free boundary conditions in the context of conformal loop ensembles.
\end{abstract}
%=========================================================================%
\section{Introduction}
Many two-dimensional critical lattice models exhibit interfaces in their configurations, whose behavior is highly informative hence desirable. The Schramm-Loewner evolution (SLE) is a mathematical theory developed around such motivation and proved to be so successful that it has become an indispensable companion to two-dimensional critical phenomena.

SLE$_\kappa$ is a one-parameter family of stochastic processes, generated through Loewner equation driven by a Brownian motion with positive diffusivity $\kappa$, and characterized by conformal invariance and the domain Markov property \cite{schramm, schlawwer,wernerrev,kager}. It is almost surely depicted by a continuous transient path in the chordal case \cite{RohdeSchramm}; a discrete version of this path is "essentially" arised by imposing Dobrushin boundary conditions in a two-dimensional critical lattice model on simply connected domains. The latter sitution can be realized as one open (boundary to boundary) interface, among possibly many closed interfaces, and in very large scales (or for very fine lattices) one can focus on that single interface to extract information from the system via SLE tools. This leads naturally to a geometric interpretation of the critical phenomena in the sense that the critical behavior can be translated to macroscopic geometrical properties of configurations. There is a long list of critical discrete models that SLE is believed (or further proved) to serve as their scaling limit; the loop-erased random walks ($\kappa=2$), self-avoiding walks ($\kappa=\frac{8}{3}$), Ising ($\kappa=3$), dimer ($\kappa=4$), and FK-Ising ($\kappa=\frac{16}{3}$) models, percolation ($\kappa=6$), uniform spanning trees ($\kappa=8$), and also families of Q-state Potts ($\kappa=\frac{4\pi}{\arccos(-\sqrt{Q}/2)}$) and O($n$) ($\kappa=\frac{4\pi}{2\pi-\arccos(-n/2)}$ for dilute phase and $\kappa=\frac{4\pi}{\arccos(-n/2)}$ for dense phase) models are all included, and any such list will not be exhaustive. The spectrum of critical exponents as well as many observables can then be calculated thanks to this insight, and even novel features of criticality show up in this way. However, once we ask about one chordal path, the question of multiple paths is almost inevitable. From one point of view, we can consider the problem as following: $2m$ points are marked on the boundary of a simply connected domain, and joined by $m$ curves. There will be $C_m=\frac{1}{m+1}{2m \choose m}$ number of planar link patterns. Considering this setup as a random system, what probability measures are likely to set the situation identical to the corresponding one in the scaling limit of a two-dimensional critical statistical mechanics model? For this to happen, a priori we infer that the paths have to close simultaneously in this scenario and no growth path would be swallowed by the region split from the whole by any other path in the pattern. For $\hspace{.1cm}m=2\hspace{.1cm}$ for example, the problem is encoded by $\kappa$ and a drift term -- a function of the cross-ratio of the four boundary points -- manifesting interaction. This is the case we are interested in. Moreover, the case of single interface can be included ($m=1$), but with treating both end points of the path symmetrically so that two SLE processes are interacting with each other. On the other hand, a primary inspiration in multiple SLEs problem may be the prediction of the relation between SLE and Dyson's Brownian motion \cite{CardyDyson}.

There have been two main approaches to construct multichordal SLEs: the local approach, which is the canonical view of SLE as a growth process, and the global approach or the configurational measure, which focuses on compatible conformally covariant partition functions indexed by (planar) connectivity patterns. For the former, the components are short time scales independence and drift terms responsible for interactions. These lead to commutation relations for multiple SLEs. The drift terms have explicit relations to some positive smooth functions called partition functions, and are fixed by conformal invariance and martingale property \cite{BBK,Dubedat}. Identically in the latter, the partition functions determine the Radon-Nikodym derivative (quantifying the absolute continuity) of global multiple SLE measures with respect to the product measures of independent SLEs, in terms of the conformally invariant Brownian loop measure \cite{BefPelWu}. If exist at all, they are expected to be the continuum version of normalized partition functions of related lattice models. Explicit formulas for pure partition functions, i.e. the normalized partition functions of each link pattern, are available in two cases of $\kappa=2$ \cite{KKP} and $\kappa=4$ \cite{PeltolaWu}. At least in the case of simple paths ($\kappa\leq4$\footnote{However there is a possibility of extending the result to $\kappa\in(4,6]$, due to the relation to discrete models \cite{BefPelWu}}), both approaches give rise to the same unique notion of multiple SLEs.

Not surprisingly, the corresponding transitions of single SLE path from simple ($0<\kappa\leq4$) to non-simple ($4<\kappa<8$), and then space-filling ($\kappa\geq8$) somehow hold in multiple cases and for the same intervals, SLE paths are almost surely simple without touching, non-simple with touching, and again space-filling, respectively. However, we should emphasize that the concept of uniqueness is more delicate in multiple cases and fails for $\kappa\geq8$ \cite{PeltolaWu}.

There is a notable connection between SLE and Conformal Field Theory (CFT) made by the loop O($n$) models -- the loop ensemble interpretation of many statistical mechanics models -- believed to be, for a certain range of parameter $n$ and at criticality, in correspondence with minimal series in CFT. Loop models is a major context in critical phenomena, also related to the mathematical theory of Conformal Loop Ensembles (CLE), a one-parameter family of countably infinite non-crossing loops, the only random loop ensemble which satisfies the conformal restriction property. It is the natural candidate for the full scaling limits of SLE-host models with the same parameter $\kappa$, defined in the range $\kappa\in(\frac{8}{3},8]$, and explored by keeping track of branches of SLE$_\kappa(\kappa-6)$ processes \cite{werself,sheffield,sheffwer}. Loop models are in turn directly related to CFT through the Coulomb gas approach \cite{nienhuis}, which makes a certain relation between $n$ and the stiffness of a random surface model. In a particular case, each conﬁguration of randomly oriented loops maps onto an instance of a height function, which gives a $c=1$ conformal field theory. The level lines of this height model with certain boundary data are identical to SLE and CLE with $\kappa=4$. This is the free bosonic field, mathematically known as the Gaussian Free Field (GFF) \cite{schsheff,sheff}, a very common object prevailing in various aspects of physics and mathematics, considered as the two-dimensional analog of the Brownian motion. The level lines of GFF is the scaling limit of the level sets of a discrete height function, which is in bijection to the dimer model on the bipartite dual lattice \cite{kenyondominoconf,kenyondomino}. They are also conjectured to be the scaling limit of the double-dimer loop ensembles \cite{kenyondouble,dubedatdeformation,basokchelkak}. 

In the language of CFT, the above-mentioned interfaces are non-local objects; one may use non-unitary representations, arised from the boundary scaling exponent $h_{1,2}$ of the Kac table, to predict the asymptotic behavior of associated correlations \cite{AA}. Furthermore, some conformal field theory arguments show that by some probabilistic presumptions, taken for granted, one can characterize multiple SLE growth processes, and so the scaling limit of likely related statistical mechanics models. This has been already done for the Ising model and percolation \cite{BBK}. Indeed, the multiple interfaces of the Ising and/or FK-Ising models have been the subject of many mathematical works \cite{Izy,Wu,KempSmir,BefPelWu,PelWu}. 
Still in analogy, the literature on multiple iterfaces is not as vast as that on single interface. One reason may be due to higher mathematical complexity; conformal invariance and the domain Markov property are not enough to characterize these interacting interfaces, and some extra properties such as boundary perturbation, and a cascade relation to express the marginal law of one path in a global setting, are have to be known to identify these sets of chordal paths \cite{KozLaw,PeltolaWu}.

In the case of $\kappa=4$, multichordal interfaces are expected to be multiple level lines of the Gaussian free field with certain alternating boundary data (the connection probabilities were shown to be the same in both cases). In this special case, the marginal law of one curve considered globally is a symmetric SLE$_4(\rho)$ process, in contrast to other values of $\kappa$ where the marginal laws of individual curves are more general variants of chordal SLE$_\kappa$. For a certain height $\lambda$ and boundary data $\lambda\backslash -\lambda\backslash \lambda\backslash -\lambda$ in GFF setting, given one path, another is the level line of GFF on the remaining domain with Dobrushin boundary data. So the conditional law of one path given another is the chordal SLE$_4$ in either case.

For global multichordal SLEs, the space of configurations is partitioned to some topologically inequivalent arch configurations. An important family of observables associated to this picture is the set of crossing or connection probabilities, (may conjecturally) related to conformally invariant scaling limit of corresponding observables in a critical statistical mechanics model, including Cardy's formula\footnote{This is in fact the only connection probability explicitly known for percolation.} as the most influential representative \cite{Cardy,Smirnov}. Pure partition functions are central objects in this respect. These observables have been studied in several works \cite{AA,BBK,KW1,KW2,Wu,PeltolaWu,PelWuCross} for some discrete models and/or multiple SLEs, and also in CLE background \cite{MillerWerner}.

\begin{figure}[h]
	\includegraphics[width=.7\textwidth]{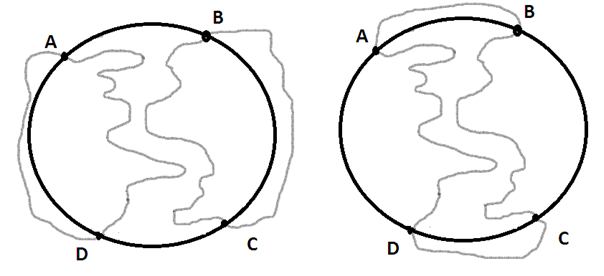}
	\centering
	\caption{A sketch of two connectivity patterns with wired/free/wired/free boundary conditions.}
	\label{Fig1}
\end{figure}

In the example of four boundary points, or equivalently local 4SLE$_\kappa$, there are only two (planar) connectivity patterns. When considered in the context of conformal loop ensembles, each SLE path can be viewed as a path going to complete a loop, part of it already explored, i.e. a wired/free/wired/free boundary conditions (Figure \ref{Fig1}). By Dub\'edat's commutation relations \cite{Dubedat,BBK} it turns out that the hook-up probability -- the probability that those wired arcs belong to the same loop -- in a rectangular domain $[0,L]\times[0,1]$ with vertical sides wired, should be of the following form 
\begin{align}
\frac{Y_\kappa(L)}{Y_\kappa(L)+\theta_\kappa Y_\kappa(1/L)}
\label{Dub}
\end{align}
for an explicit function $Y_\kappa(L)$ and some value $\theta_\kappa$ \cite{MillerWerner,Dubedat}. A genuine computation (without referring to the conjectural relations to discrete models) of this probability for CLE$_\kappa$ in conformal squares, $\kappa\in(\frac{8}{3},8)$, has been done, which gives $\theta_\kappa=-2\cos(4\pi/\kappa)$, that is equal to the parameter $n$ of the related O($n$) model. This further supports these conjectural relations predicted by the Coulomb gas methods. There is also a relation $\frac{2}{\kappa}+\frac{2}{\tilde{\kappa}}=1$, uncorrelated with the celebrated duality relation of SLE $\hspace{.1cm}\kappa\kappa'=16$, which relates $\kappa\in(\frac{8}{3},4)$ and $\tilde{\kappa}\in(4,8)$ with the same connection probabilities \cite{MillerWerner}. In a lattice model, such as the double-dimer model, the above setup can be considered as wiring two outer arcs of the boundary by dimers, which alternately belong to one of the contributing dimer models. Connection probabilities are defined then. From now on, we will concern ourselves with this problem.

Computing some loop-related quantities, one can do identification between (proved or expected) scaling limits of discrete models and the parameter $\kappa$ of SLE/CLE. Here we want to compute an observable, expected to be in close relation to 4SLE$_\kappa$ ($\kappa=4$) processes, the hook-up probability for the double-dimer model, taking advantage of the Grassmannian representation of the dimer model, the same tool and very similar approach that we employed in the previous work \cite{NS} to compute some loop-based observables in the double-dimer model. After a very brief review on the dimer and the double-dimer models, we introduce the Grassmannian representation of the dimer model, an effective method we exploit to compute symmetric and pure partition functions of the double-dimer model on a rectangular domain of the square lattice with four marked points (monomers) on the corners. We then obtain their asymptotic behavior in the continuum, where, with the aid of some series identities related to complete elliptic integrals, we show that the connection probabilities are consistent with the known results in the upper half-plane. This supports the strong, yet unproven, prediction of the 
relation between the double-dimer loop ensembles and the level lines of the Gaussian free field and conformal loop ensembles with $\kappa=4$. As almost always, the notion of bipartite lattice is crucial in solutions.

There are some link patterns in the general case of $2m$ marked boundary points, the so-called rainbow link patterns or self-surrounding arch configurations, where each next arch encloses the previous one. These link patterns are particularly interesting because the conformal blocks and the pure partition functions are equal in these cases, the condition which does not hold in general \cite{BBK,PeltolaWu}. The approach we follow is also quite straightforward in the case of these patterns in the double-dimer model. We will comment on this at the end of section \ref{connection}.

In \cite{KW1} and \cite{KW2} formulas for connection probabilities of some discrete models, including the double-dimer model, were derived which cover the special case of four boundary points. However, although this case has already been obtained rigorously in the upper half-plane, we think our primarily Grassmannian approach is interesting by itself and the result obtained here is still beneficial because explicit formulas for Jacobi elliptic functions (inverse Schwarz-Christoffel transformations) are not available in general, and they are computed only through complicated numerical methods.

To be self-contained while not interrupting the main discussion, two short appendices, one on some formulas related to complete elliptic integrals and the other on some series identities, have been added. 

%=========================================================================%
\section{Preliminaries}
In this section, we introduce the dimer model and its Grassmannian representation, the powerful tool that we use to compute a loop-related observable in the double-dimer model. The approach is very much like one in \cite{NS}; the reader may consult this paper as well as the original reference \cite{HaynPlechko}, to see the derivation of identities brought here and on Grassmannian analysis of the close-packed dimer model.

Given a connected simple planar graph, a dimer covering or perfect matching of the graph is a collection of edges of the graph such that every vertex is visited exactly once. One can consider fugacities for these edges, or rather dimers, so the weight of each dimer configuration is defined as the multiplication of fugacities of contributing edges, and the partition function is then the sum of all these weights on all dimer configurations \cite{kenyonintro}.

The two-dimensional dimer model is an exactly solvable model, playing an important role in many areas of interest including statistical mechanics, condensed matter physics, combinatorics, random matrix theory, partition theory, and many more. It was primarily proposed to simulate the adsorption of diatomic molecules on two-dimensional surfaces \cite{fowlerrush}, but did not gain much attention from empirical studies, as in its close-packed case it is too abstract, yet solvable. It is commonly known in mathematics as perfect matchings, drawing on combinatorial methods, even implicitly, in solving dimer problems. The original solution of the close-packed dimer model goes back to the early '60s, and was given by Kasteleyn \cite{Kasteleyn}, and independently by Temperley and Fisher \cite{TemperleyFisher}. It is in fact the Pfaffian (the square root of the determinant) of a signed adjacency matrix, called the Kasteleyn matrix.

The first question about the dimer model is, how many dimer configurations exist for a given graph? This number is equal to the partition function of the dimer model with trivial fugacities 1, which is the uniform dimer model.

In general, some monomers can be present in the model, making the model more realistic but intractable analytically. Nevertheless, there are rare cases, particularly when monomers only exist on the boundary, that the problem submits itself to an exact solution. This particular case pertains to our discussion in the sequel. 

The practical use of the dimer model may mainly be made through its close relation to other statistical mechanics models, offering new insight and effective methods in dealing with other models such as the Ising model, vertex models and spanning trees \cite{kasteleynising,fisher,kenmilshefwil}. The fermionic nature of the dimer model is in the heart of most of these relations and further supplies connections to determinants and combinatorial methods to tackle some tough models to analyze \cite{Baxter}.

The dimer model on bipartite lattices is of especial significance because of the notion of height functions and rough random surfaces, where the Gaussian free field comes into play establishing the relation between the scaling limit of level sets of the dimer height function and conformal loop ensembles with $\kappa=4$ \cite{kenyondominoconf,kenyondomino}.

An alternative (but similar in spirit) way to give a loop-description of dimer configurations \cite{kenyondouble} is to put two dimer configurations on each other. Other than some individual (double) dimers, there will be a collection of simple loops, each of which is made in two ways. So the weight of each uniform double-dimer configuration, i.e. where both contributing dimer models are uniform, is then proportional to $2^l$, where $l$ is the number of (non-zero area) double-dimer loops in the configuration. Here, there is no interaction; we are only interested in the geometric picture. The partition function is then the product of the partition functions of the two contributing dimer models. Typically we are interested in the statistics of very long loops in very large lattices, that is to say macroscopic loops in the (approximate) continuum of the model. Several remarkable related results have been found so far \cite{kenyondouble,dubedatdeformation,basokchelkak}.

There are various solutions to the dimer problem based on combinatorial analyses and transfer matrix methods. Grassmann techniques are also helpfully available \cite{Samuel,HaynPlechko}, thanks to the fermionic features of the dimer model substantiated in its tight relation to the Ising model \cite{kasteleynising,fisher}. Generally these anti-commuting variable analyses of the problems have the advantage of being concise and more to the point, and avoid lengthy solutions because of making direct connection to determinantal arguments. We will make use of these opportunities in our work.

For a square sublattice of size $M\times N$, $M$ and/or $N$ even, the partition function of the uniform dimer model can be written as
\begin{align}
\mathcal{Q}_0=\int \prod_{n=1}^{N}\prod_{m=1}^{M} \dd\eta_{m,n} (1+\eta_{m,n}\eta_{m+1,n}) (1+\eta_{m,n}\eta_{m,n+1})
\label{nilpartition}
\end{align}
with the free boundary conditions: $\eta_{M+1,n}=\eta_{m,N+1}=0$, where $\eta_{m,n}$ are commuting nilpotent variables attached to every site, satisfying $\eta_{m,n}^2=0$, $\int\dd\eta_{m,n}\eta_{m,n}=1$ and $\int\dd\eta_{m,n}=0$, to ensure that each site is met by a dimer exactly once. The zero index indicates the close-packed dimer model without any defect (monomer) in the system. The representation \eqref{nilpartition} is equivalent to the hafnian of the adjacency matrix of the underlying graph (here the square lattice).

A huge step forward in making the problem tractable is to relate the representation \eqref{nilpartition} to a fermionic (Grassmannian) form, i.e. a determinant. This corresponds to Kasteleyn's combinatorial solution to the close-packed dimer problem; one can introduce an appropriate orientation in a planar lattice, such that the partition function of the dimer model is equivalent to a Pfaffian. Thanks to the mirror symmetry available in two dimensions \cite{HaynPlechko}, one can bring together all variables with the same indices to integrate, and eventually replace the commuting nilpotent variables $\{\eta_{m,n}\}$ with anti-commuting ones $\{c_{m,n}\}$, so \eqref{nilpartition} becomes
\begin{align}
\mathcal{Q}_0=\int\prod_{n=1}^{N}\prod_{m=1}^{M}\overset{m}{\overrightarrow{\dd c_{m,n}}}\  \exp\left\{\sum_{m=1}^{M}\sum_{n=1}^{N} [c_{m+1,n}c_{m,n}+(-1)^{m+1}c_{m,n+1}c_{m,n}]\right\}
\label{grasspartition}
\end{align}
\begin{figure}[h]
	\includegraphics[width=.8\textwidth]{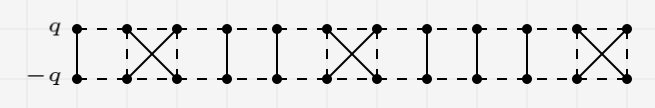}
	\centering
	\caption{An instance of admissible transformed dimer configuration on coupled Fourier modes $q$ and $-q$, for a square lattice with $M=12$. Each cross-form has weight 1, and each vertical bond has weight $2\cos\frac{\pi q}{N+1}$.}
	\label{Fig2}
\end{figure}
\begin{figure}[h]
	\includegraphics[width=.8\textwidth]{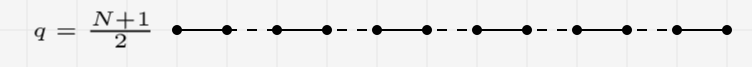}
	\centering
	\caption{The line indicating the Fourier mode $q=\frac{N+1}{2}$. Each horizontal bond has weight $1$.}
	\label{Fig3}
\end{figure}
along with the fermionic action $\mathcal{S}_0$ of the close-packed dimer model
\begin{align}
-\mathcal{S}_0=\sum_{m=1}^{M}\sum_{n=1}^{N} [c_{m+1,n}c_{m,n}+(-1)^{m+1}c_{m,n+1}c_{m,n}]
\label{actionpure}
\end{align}

The Gaussian representation \eqref{grasspartition} is homogeneous enough to be computed. We do Fourier transformation only in one direction, which for our purposes gives rise to more suitable results
\begin{align}
c_{m,n}=\sqrt{\frac{2}{N+1}} i^{n}\sum_{q=1}^{N}c_{m,q}\sin\frac{\pi qn}{N+1}.
\label{fourier}
\end{align}

We should be careful about the signs of the terms with respect to each other in partition functions expansions, otherwise only the overall absolute value is significant. Using \eqref{fourier}, the action \eqref{actionpure} becomes
\begin{align}
-\mathcal{S}_0=\sum_{m=1}^{M}\sum_{q=1}^N[-c_{m+1,q}c_{m,-q}-i(-1)^{m+1}\cos\frac{\pi q}{N+1}c_{m,q}c_{m,-q}]
\end{align}
implies decoupling of lattice connections to some strips coupling two Fourier modes $q$ and $-q\equiv N+1-q$ (Figure \ref{Fig2}) with the weights $1$ and $2\cos\frac{\pi q}{N+1}$ for every cross-form and vertical bond respectively, and a line instead of one strip indicating the mode $q=\frac{N+1}{2}\hspace{.1cm}$ when $N$ is odd (Figure \ref{Fig3}), with the weight $1$ for every horizontal bond. Now with the aid of some elementary combinatorics \cite{NS} we can compute \eqref{grasspartition} which is
\begin{align}
\mathcal{Q}_0=\prod_{q=1}^\frac{N}{2}\left[\sum_{p=0}^\frac{M}{2}\left( \begin{array}{c} M-p \\ p \end{array} \right)(2\cos\frac{\pi q}{N+1})^{M-2p}\right]
\label{freepartitione}
\end{align}
for $N$ even, and replacing the upper bound of the product by $\frac{N-1}{2}$ for $N$ odd. This is in agreement with the previously known result \cite{Kasteleyn,TemperleyFisher}. The partition function of the uniform double-dimer model is simply $\mathcal{Z}_0=\mathcal{Q}_0^2$.

We can also write \eqref{freepartitione} in the following more concise form 
\begin{align}
\mathcal{Q}_0=\prod_{q=1}^\frac{N-1}{2}\left[(-1)^\frac{M}{2}U_M(ix_q)\right]\hspace{.3cm}(N\hspace{.1cm}\text{odd})
\label{chebyshevpartition}
\end{align}
where $x_q=\cos(\frac{\pi q}{N+1})$, and $U_M(x)$ is the Chebyshev polynomial of the second kind (see Appendix B of \cite{NS}).
\begin{figure}[h]
	\includegraphics[width=.8\textwidth]{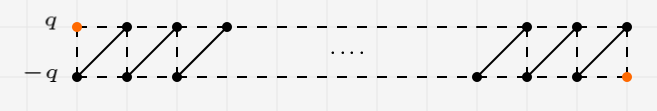}
	\centering
	\caption{An instance of coupled Fourier modes $q$ and $-q$ with two monomers (in color) sticking to its ends.}
	\label{Fig4}
\end{figure}
\begin{figure}[h]
	\includegraphics[width=.8\textwidth]{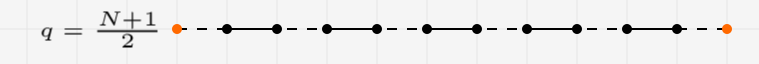}
	\centering
	\caption{The Fourier mode $q=\frac{N+1}{2}$ with two monomers (in color) sticking to its ends.}
	\label{Fig5}
\end{figure}

In a similar way, we can compute the corresponding monomer-dimer partition function when there are some monomers on the boundary. A simple situation relevant to the forthcoming section is when there are two monomers on the opposite sides of the rectangle, for example at $(1,n_1)$ and $(M,n_2)$. If we assume $M$ is even, both $n_1$ and $n_2$ have to be of the same parity so that the associated monomer-dimer partition function is non-zero. The presence of the monomers can be fulfilled by insertion of $c_{1,n_1}c_{M,n_2}$ in front of every term in the expansion of the partition function of the close-packed dimer model. After doing Fourier transformation \eqref{fourier}, this leads to one strip bearing two monomers at its ends (Figure \ref{Fig4}), others as before (Figure \ref{Fig2}). In the case $N$ is odd, there will be another possibility of two monomers on $q=\frac{N+1}{2}$ line (Figure \ref{Fig5}). The corresponding partition function turns out to be
\begin{align}
\mathcal{Q}_{n_1,n_2}=\frac{2}{N+1}\left((-1)^\frac{n_1+n_2}{2}\sum_{q=1}^{\frac{N-1}{2}}2\frac{\sin\frac{\pi qn_1}{N+1}\sin\frac{\pi qn_2}{N+1}}{U_M(q)}-1\right)\mathcal{Q}_0
\label{monomerpartition}
\end{align}
where $\mathcal{Q}_0$ is \eqref{chebyshevpartition}, $U_M(q)\equiv |U_M(i\cos\frac{\pi q}{N+1})|$, and the index $n_1,n_2$ shows the vertical indices of the monomers. The next step is to use the partition function \eqref{monomerpartition} in our setup for a conditioned version of the double-dimer model with wired/free/wired/free boundary conditions on four sides of the rectangle $M\times N$.

%=========================================================================%
\section{Connection probabilities}
\label{connection}
A key family of observables readily defined on global multichordal SLEs is the crossing or connection probabilities for SLE paths emanating from some (even number) marked points on the boundary of a simply connected planar domain. With the situation interpreted in loop ensemble setting, it is comparable with the corresponding probabilities for discrete O($n$) models, which implies that if such a model has a conformally invariant scaling limit, this should be CLE with the appropriate $\kappa$. 

In the case of four marked boundary points (so two connecting paths) there are only two possible topological connectivities; we are going to derive the so-called hook-up probability in the correspnding situation in the double-dimer model on a rectangle. An observable relevant to the simplest case of single chordal SLE is the left-passage probability, which we computed for the double-dimer model on a rectangle using very similar method to the present one \cite{NS}.\vspace{.3cm}

\begin{figure}[h]
	\includegraphics[width=1\textwidth]{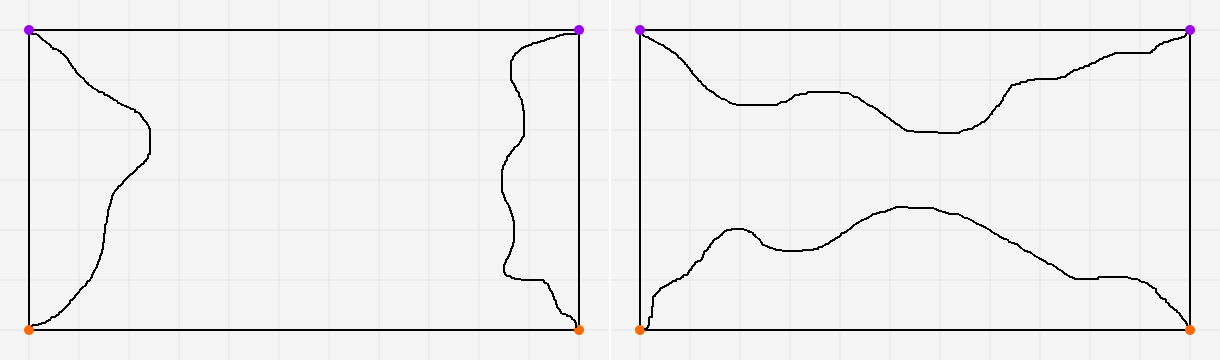}
	\centering
	\caption{A sketch of Type I (left) and Type II (right) double-dimer configurations. The monomers in either dimer model are in the same color.}
	\label{Fig6}
\end{figure}

%=============================================================================%
\textbf{\large The discrete setting.} Suppose we insert two pairs of monomers on the corners of a rectangular $M\times N$-subdomain of the square lattice, each pair in one dimer model horizontally inline. In every double-dimer configuration, there will be a pair of paths from one corner to another, yielding two distinct arch-types \cite{BBK}, or two types of configurations (Figure \ref{Fig6}): Type I and Type II, where each pair joins two vertically and horizontally inline monomers, respectively. To realize both types, $N$ should be odd in this setting (so $M$ is even). 

We need to distinguish between these two types of configurations. Consider we swap the position of right monomers in the dimer models, up to down and vice versa. This will cause desirable nonlocal effects in double-dimer configurations. Type II configurations will disappear. Type I configurations will nevertheless remain unchanged. This is because we can consistently identify the membership of which dimer model, for each dimer contributed in those two paths, due to the bipartiteness of the square lattice. Swapping the positions of monomers suppress any path joining two horizontally inline monomers, while being only a flip in the ownership of dimers (to one dimer model or the other), along any path joining two vertically inline monomers.

Let's denote by $\mathcal{Z}_I$ and $\mathcal{Z}_{II}$, and $\mathcal{Z}=\mathcal{Z}_I+\mathcal{Z}_{II}$, the pure partition functions of Type I and Type II, and the symmetric partition function of the double-dimer model respectively, all normalized by the partition function of the uniform double-dimer model, in agreement with the terminology used for multiple SLEs. By considering \eqref{monomerpartition}, once for $n_1=n_2=1$ and $n_1=n_2=N$, and then $n_1=1,n_2=N$ and $n_1=N,n_2=1$, we obtain the following
\begin{align}
\mathcal{Z}=\left(\frac{2}{N+1}\left(\sum_{q=1}^{\frac{N-1}{2}}2\frac{\sin^2\frac{\pi q}{N+1}}{U_M(q)}+1\right)\right)^2
\label{Z}
\end{align}
and
\begin{align}
\mathcal{Z}_I=\left(\frac{2}{N+1}\left((-1)^\frac{N+1}{2}\sum_{q=1}^{\frac{N-1}{2}}2\frac{\sin^2\frac{\pi q}{N+1}(-1)^{q+1}}{U_M(q)}-1\right)\right)^2
\label{ZI}
\end{align}
where $\mathcal{Z}=\mathcal{Q}_{1,1}\mathcal{Q}_{N,N}=\mathcal{Q}_{1,1}^2$ and $\mathcal{Z}_I=\mathcal{Q}_{1,N}\mathcal{Q}_{N,1}=\mathcal{Q}_{1,N}^2$.

\begin{figure}[h]
	\includegraphics[width=.8\textwidth]{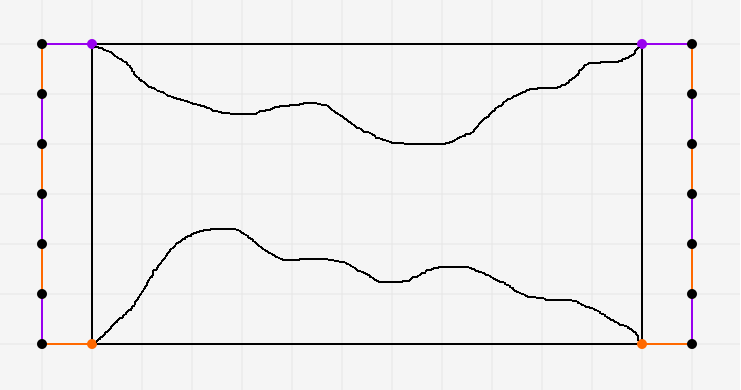}
	\centering
	\caption{A schematic picture of Type II double-dimer configuration on a $(M+2)\times N$ rectangle with wired/free/wired/free boundary conditions. There are two double-dimer paths making one loop (together with the wired boundary arcs), the so-called hook-up event. The monomers in either dimer model are in the same color and in agreement with the ownership of dimers on the boundary.}
	\label{Fig7}
\end{figure}

Indeed the above picture can be interpreted as a double-dimer model with alternating boundary conditions wired/free/wired/free. Assuming wired (with dimers alternately belong to one dimer model and the other) boundary conditions, for example on the remaining two boundary arcs of a rectangular $(M+2)\times N$ domain where the original $M\times N$ rectangle is removed, each double-dimer path from one monomer to another completes a loop, part of it already prescribed (Figure \ref{Fig7}). The aforementioned possibilities are: Type I configurations, where there are two separate loops, each meets two vertically inline monomers, and Type II where all four monomers belong to the same loop. We call the pure state of Type II the hook-up event in accordance with the terminology used in \cite{MillerWerner}. In the scaling limit, this can be compared with a 4SLE event or an event associated with a conditioned version of CLE\footnote{Defining a wired CLE may require more care and there are delicacies concerning boundary conditions in the dimer model. However, that kind of methods like the one applied here is not suitable to deal with such issues.}. Following this new interpretation, the new symmetric double-dimer partition function is 
\begin{align}
\mathcal{Z}_{dd}=4\mathcal{Z}_I+2(\mathcal{Z}-\mathcal{Z}_I)
\end{align}
where we have considered the additional contribuion of two loops in Type I configurations and likewise one loop in Type II. So the hook-up probability becomes
\begin{align}
\mathbb{P}(\text{Type II})&=\frac{2(\mathcal{Z}-\mathcal{Z}_I)}{\mathcal{Z}_{dd}}\nonumber\\
&=\frac{\mathcal{Z}-\mathcal{Z}_I}{\mathcal{Z}+\mathcal{Z}_I}
\label{hookup}
\end{align}
\vspace{.1cm}
%=============================================================================%

\textbf{\large The continuum limit.} We want to know about the asymptotic behavior of \eqref{Z} and \eqref{ZI}, mainly to find the scaling limit of the hook-up probability \eqref{hookup}, for a $[0,L]\times[0,1]$ rectangle
\begin{align}
H(L)=\frac{Z(L)-Z_I(L)}{Z(L)+Z_I(L)}
\label{hookupL}
\end{align}
where by $Z(L)$ we mean the scale invariant quantity $Z(L)\equiv (MN)\mathcal{Z}$, and so on. 

On the other hand, commutation relations for $\text{SLE}_\kappa$ \cite{Dubedat} show that the corresponding probability $H(L)$ for CLE should be of the form \eqref{Dub}, for an explicit function $Y_\kappa(L)$ and some value $\theta_\kappa$, proved to be equal to $-2\cos(4\pi/\kappa)$ \cite{MillerWerner}. This is the parameter $n$ of the discrete $O(n)$ model, which conjecturally converges to CLE$_\kappa$. This implies that for the double-dimer model ($n=2$), which is expected to have CLE$_4$ as the scaling limit, the following should hold
\begin{align}
Z(1/L)-Z_I(1/L)&=Z_I(L)\nonumber\\
&=Y(1/L)
\label{symmetry}
\end{align}
where we have called $Y(L)\equiv Z(L)-Z_I(L)$, in harmony with \eqref{Dub}.

Though by symmetry it has to happen that $Z_I(L)=Y(1/L)$, we show it explicitly to make sure of self-consistency.   

Again, the argument is quite similar to that in \cite{NS}. For very large $M$ and $N$, one may note that the concentration of \eqref{Z} and \eqref{ZI} is on $q$-modes when $\cos\frac{\pi q}{N+1}\approx 0$. An expansion near $\frac{\pi}{2}$ leads to $U_M(q)\approx\cosh(\frac{M}{N}\pi k)$, for small $\frac{k}{N+1}$ and large $M$, and $q=\frac{N+1}{2}-k$. So \eqref{Z} and \eqref{ZI} are asymptotically
\begin{align}
Z(L)=4L\left(\sum_{k=1}^{\infty}\frac{2}{\cosh(L\pi k)}+1\right)^2,
\label{Zasymptot}
\end{align}
\begin{align}
Z_I(L)=4L\left(\sum_{k=1}^{\infty}\frac{2(-1)^{k+1}}{\cosh(L\pi k)}-1\right)^2
\label{ZIasymptot}
\end{align}

Now the explicit expression of $Y(L)$ is
\begin{align}
Y(L)=4L\left(\sum_k\frac{4}{\cosh(L\pi(2k-1))}\right)\left(\sum_k\frac{4}{\cosh(L\pi(2k))}+2\right) 
\label{Y}
\end{align}

First we rewrite \eqref{ZIasymptot} and \eqref{Y} as
\begin{align}
Z_I(L)=4L\left(\sum_{k_1,k_2}\frac{8(-1)^{k_1+k_2}}{\cosh(L\pi(k_1+k_2))+\cosh(L\pi(k_1-k_2))}-\sum_k\frac{4(-1)^{k+1}}{\cosh(L\pi k)}+1\right)
\label{ZIrewrite}
\end{align}
and
\begin{align}
Y(L)=4L\left(\sum_{k_1,k_2}\frac{32}{\cosh(L\pi(2(k_1+k_2)-1))+\cosh(L\pi(2(k_1-k_2)-1))}+\sum_k\frac{8}{\cosh(L\pi(2k-1))}\right)
\label{Yrewrite}
\end{align}

Then we sum over "$k_1-k_2$" in the first summation, using the identities \eqref{cosh2} and \eqref{cosh4}, so \eqref{ZIrewrite} and \eqref{Yrewrite} become
\begin{align}
Z_I(L)=4L\left(\sum_{k=1}^\infty\frac{4(-1)^k k}{\sinh(L\pi k)}+\sum_{k=1}^\infty\frac{4}{1+\cosh(2L\pi k)}+1\right)
\label{ZIfinal}
\end{align}
\begin{align}
Y(L)=4L\sum_{k=1}^\infty\frac{8(2k-1) }{\sinh(L\pi(2k-1))}
\label{Yfinal}
\end{align}
If we consider the following result of Cauchy's residue theorem
\begin{align}
\text{pr.v.}\int_{-\infty}^{\infty}f(x)\dd x=2\pi i\sum\text{Res}f(z)+\pi i\sum\text{Res}f(z)
\label{residue}
\end{align}
for the complex function $f(z)=\frac{z}{\sin( z)\sinh(\alpha z)}$, where the fist sum extends over all poles in the upper half-plane and the second over all simple poles
on the real axis, we obtain the following equality
\begin{align}
\sum_{k=-\infty}^\infty\frac{k(-1)^k}{\sinh(\alpha k)}+2\sum_{k=1}^\infty\frac{k(-1)^k}{\alpha^2\sinh(k/\alpha)}=0
\end{align}
Using this and the identity \eqref{sumsinh}, we come to the intended symmetry \eqref{symmetry}.

We now argue, via Schwarz-Christoffel transformations, that the upper half-plane version of \eqref{hookupL} is equal to what obtained previously in \cite{KW1,KW2}, and is indeed identical to the corresponding known results for multiple SLE$_4$ or conditioned CLE$_4$ in the upper half-plane with four marked points on the real axis. Such evidence supports the prediction of CLE$_4$ being the scaling limit of double-dimer loop ensembles. First we note that the cross-ratio $x$ of $(\infty,0,1-x,1)$ in the upper half-plane and the aspect-ratio $L$ of the (conformal) rectangle $(0,L)\times(0,1)$ are related by \cite{MillerWerner}
\begin{align}
L=\frac{F(\frac{1}{2},\frac{1}{2},1;1-k^2)}{2F(\frac{1}{2},\frac{1}{2},1;k^2)}
\end{align}
where $x=(\frac{1-k}{1+k})^2$, and $F$ is the hypergeometric function $\,_2F_1$. To express $H(L)$ as a function of $x$, we use a series identity \eqref{serieselliptic}, which relates $Y(L)$ and complete elliptic integrals of the first kind $K(k)$ \eqref{ellipticf}
\begin{align}
&Y(L)=\frac{k_1^2}{\pi^2}K(k_1)K'(k_1),\hspace{1.37cm}k_1=\frac{2\sqrt{k}}{1+k}\\
&Y(1/L)=\frac{k_2^2}{\pi^2}K(k_2)K'(k_2),\hspace{1cm}k_2=\frac{1-k}{1+k}
\label{YK12}
\end{align}

Here $K(k_1)$ is proportional (with proportionality constant $\frac{\pi}{2}$) to the hypergeometric function $\,_2F_1(\frac{1}{2},\frac{1}{2},1;k_1^2)$, and is in fact a conformal mapping from the upper half-plane to a rectangle with aspect-ratio $L$ and prevertices $(-\frac{1}{k_1},-1,1,\frac{1}{k_1})$. The functions in \eqref{YK12} can be rewritten in terms of $k$, with the aid of \eqref{ellipticrelation}
\begin{align}
&Y(L)=\frac{2k}{\pi^2}K(k)K'(k),\nonumber\\
&Y(1/L)=\frac{(1-k)^2}{2\pi^2}K(k)K'(k).
\label{YK}
\end{align}
Summing up all these, we arrive at the following expression for $H(L)$ \eqref{hookupL}
\begin{align}
H(L)&=\frac{2k}{1+k^2}\nonumber\\
    &=\frac{1-x}{1+x}
\label{final}
\end{align}
concerning the relation between $k$ and $x$. The probability \eqref{final} is equal to the corresponding one obtaind for CLE$_4$ according to \cite{BBK,MillerWerner}.

A generalization of the above setup can be made to obtain the symmetric partition function in the case of $2m$ marked points on the boundary, that is when there are $C_m=\frac{1}{m+1}{2m \choose m}$ number of topologically inequivalent (planar) link patterns. In fact for every $2m$ marked boundary points, there is an arrangement of ownership for monomers (to one dimer model or the other) at these points, such that the partition function of the resultant double-dimer model gives the whole collection of connection configurations. To cofirm this we can begin from an arbitrary adjacent pair and determine their ownership in consistent with each other, by which we mean their ownership allows a double-dimer path to connect them so that any related addmissible sub-pattern can be realized by this choice of ownership. Then we determine the ownership of the next two immediate neighbors consistent with them and with each other, and so on. At each time the problem is similar to a four boundary points situation. In this way, the ownership of all monomers will be determined by induction. We can see that such an arrangement is unique up to an overall flip in the ownership. Now assume we alter the ownership in this setup for symmetric partition function, so that at each step of the above procedure, the ownership of the next two neighbors is exchanged (between the two dimer models) or remains unchanged, alternately. The only link pattern adapted to this new setting is the rainbow link pattern, whose most internal pair (considered in the upper half-plane) is the pair at the first step in the procedure. As such, the pure partition functions of rainbow link patterns (any cyclic permutation of them) are immediate, which is the spacial case considered in \cite{KozLaw} and also mentioned in \cite{BBK} with the terminology self-surrounding arches. For link patterns in general, all pure partition functions have been found particularly for the double-dimer model \cite{KW1,KW2}, which we will address in a forthcoming paper.
%========================================================================================%
\begin{appendices}
\section{Some formulas related to complete elliptic integrals}
\label{appendixA}
Schwarz-Christoffel transformations equip us with explicit formulas for conformal maps from the upper half-plane to the interior of any polygon; their existence are already guaranteed by Riemann mapping theorem. In the simplest cases, they boil down to elliptic integrals; for a rectangle with prevertices $(-a,-1,1,a)$, the function $f(z)=\int_0^z\frac{1}{\sqrt{(w^2-1)(w^2-a^2)}}\dd w$ works, which transforms to
\begin{align}
K(k)=\int_0^1\frac{1}{\sqrt{(1-w^2)(1-k^2w^2)}}\dd w
\label{ellipticf}
\end{align}
that is the complete elliptic integral of the first kind with modulus $k=\frac{1}{a}$. As in the situation we have studied the above reflection symmetry between prevertices does not hold, the following equality is particularly useful \cite{wolfram}
\begin{align}
\frac{K'(k)}{K(k)}=2\frac{K'(\frac{2\sqrt{k}}{1+k})}{K(\frac{2\sqrt{k}}{1+k})}=\frac{1}{2}\frac{K'(\frac{1-k'}{1+k'})}{K(\frac{1-k'}{1+k'})}
\label{ellipticrelation}
\end{align}	
where $K'(k)\equiv K(\sqrt{1-k^2})$.

There are some interesting series identities related to elliptic integrals. One we have used in section \ref{connection} is
\begin{align}
\sum_{n=1}^\infty\frac{2n-1}{\sinh(\pi(2n-1)\xi)}=\frac{k^2}{\pi^2}K^2(k)
\label{serieselliptic}
\end{align}
which is directly obtained from two other identities (relations (5.3.4.2) and (5.3.4.3) in \cite{Prudnikov})
\begin{align*}
&\sum_{n=1}^\infty\frac{n}{\sinh(\pi n\xi)}=\frac{K(k)}{\pi^2}[K(k)-E(k)],\\
&\sum_{n=1}^\infty\frac{(-1)^{n-1}n}{\sinh(\pi n\xi)}=\frac{K(k)}{\pi^2}[E(k)-(1-k^2)K(k)]
\end{align*}
where $\xi\equiv\xi(k)=\frac{K'(k)}{K(k)}$, and $E(k)$ is the complete elliptic integral of the second kind
\begin{align}
E(k)=\int_0^1\sqrt{\frac{1-k^2w^2}{1-w^2}}\dd w.
\label{elliptics}
\end{align}
\vspace{.01cm}

\section{Some series identities}
\label{appendixB}
There are many useful identities derived from the Poisson summation formula
\begin{align}
\sum_{n\in\mathbb{Z}}f(n)=\sum_{n\in\mathbb{Z}}\hat{f}(n)
\label{poisson}
\end{align}
where $\hat{f}(\xi)=\int_{-\infty}^{\infty}f(x)e^{-2\pi ix\xi}\dd x$ is the Fourier transform of $f$. We have used a few of them 
\begin{align}
\sum_{n=-\infty}^{\infty}\frac{1}{\cosh(\alpha (2n-1))+\cosh(\alpha m)}=\frac{m}{\sinh(\alpha m)},
\label{cosh2}
\end{align}
\begin{align}
\sum_{n=-\infty}^{\infty}\frac{1}{\cosh(\alpha (4n-1))+\cosh(\alpha m)}=\frac{\frac{1}{2}m}{\sinh(\alpha m)},
\label{cosh4}
\end{align}
where $m$ is a non-zero integer. Both \eqref{cosh2} and \eqref{cosh4} can be seen as special cases of the following
\begin{align}
\sum_{n=-\infty}^{\infty}\frac{1}{\cosh(\alpha n)+\cosh(\alpha m)}=\frac{2m}{\sinh(\alpha m)},
\label{cosh1}
\end{align}
which was obtained in \cite{NS} (see its Appendix B). 

For example to derive \eqref{cosh4}, consider the identity \eqref{poisson} for the function $f(x)=\frac{1}{\cosh(\alpha (4x-1))+\cosh(\alpha m)}$, which has the required conditions \cite{complex}. The RHS of \eqref{poisson} will be a sum on $\hat{f}(\xi)=\frac{\pi}{2\alpha \sinh(\alpha m)}\frac{e^\frac{-i\pi\xi}{2}\sin(\frac{\pi m}{2}\xi)}{\sinh(\frac{\pi^2}{2\alpha}\xi)}$, whose contribution for either even or odd terms equals zero, except for $\xi=0$, which is the RHS of \eqref{cosh4}.

If we repeat the above calculation for $f(x)=\frac{1}{\cosh(\alpha n)+\cosh(\alpha m)}$, then $\hat{f}(\xi)=\frac{2\pi}{\alpha \sinh(\alpha m)}\frac{\sin(2\pi m\xi)}{\sinh(\frac{2\pi^2}{\alpha}\xi)}$, and by taking $m\rightarrow 0$, we come to another identity, which we used in section \ref{connection}
\begin{align}
\sum_{n=-\infty}^{\infty}\frac{1}{1+\cosh(\alpha n)}=\sum_{n=-\infty}^{\infty}\frac{4n\pi^2}{\alpha^2\sinh(\frac{2\pi^2 n}{\alpha})}
\label{sumsinh}
\end{align}
 
\end{appendices}	
%========================================================================================%

%========================================================================================%
\end{document}